\documentclass[aps,apl,showpacs]{revtex4}

\usepackage{graphicx}

\begin{document}

\draft

\preprint{T. Unuma \textit{et al.}, JAP}

\title{
Intersubband absorption linewidth in GaAs quantum wells due to scattering by interface roughness, phonons, alloy disorder, and impurities\footnote[1]{submitted to J. Appl. Phys.}
}

\author{
Takeya Unuma\footnote[2]
{Electronic mail: unuma@issp.u-tokyo.ac.jp}$^{)}$ 
and Masahiro Yoshita
}

\affiliation{
Institute for Solid State Physics, University of Tokyo, 
5-1-5 Kashiwanoha, Kashiwa, Chiba 277-8581, Japan
}

\author{Takeshi Noda and Hiroyuki Sakaki}
\affiliation{
Institute of Industrial Science, University of Tokyo, 
4-6-1 Komaba, Meguro-ku, Tokyo 153-8505, Japan
}

\author{Hidefumi Akiyama}
\affiliation{
Institute for Solid State Physics, University of Tokyo, 
\,5-1-5 Kashiwanoha, Kashiwa, Chiba 277-8581, Japan\,
}

\date{9 Aug. 2002}

\begin{abstract}
We calculate the intersubband absorption linewidth $2\Gamma_{\mathrm{op}}$ in quantum wells (QWs) due to scattering by interface roughness, LO phonons, LA phonons, alloy disorder, and ionized impurities, and compare it with the transport energy broadening $2\Gamma_{\mathrm{tr}} = 2\hbar/\tau_{\mathrm{tr}}$, which corresponds to the transport relaxation time $\tau_{\mathrm{tr}}$ related to the electron mobility $\mu$. 
Numerical calculations for GaAs QWs clarify the different contributions of each individual scattering mechanism to the absorption linewidth $2\Gamma_{\mathrm{op}}$ and transport broadening $2\Gamma_{\mathrm{tr}}$. 

Interface roughness scattering contributes about an order of magnitude more to the linewidth $2\Gamma_{\mathrm{op}}$ than to the transport broadening $2\Gamma_{\mathrm{tr}}$, because the contribution from the intrasubband scattering in the first excited subband is much larger than that in the ground subband. 
On the other hand, LO phonon scattering (at room temperature) and ionized impurity scattering contribute much less to the linewidth $2\Gamma_{\mathrm{op}}$ than to the transport broadening $2\Gamma_{\mathrm{tr}}$. 
LA phonon scattering makes comparable contributions to the linewidth $2\Gamma_{\mathrm{op}}$ and transport broadening $2\Gamma_{\mathrm{tr}}$, and so does alloy disorder scattering. 

The combination of these contributions with significantly different characteristics makes the absolute values of the linewidth $2\Gamma_{\mathrm{op}}$ and transport broadening $2\Gamma_{\mathrm{tr}}$ very different, and leads to the apparent lack of correlation between them when a parameter, such as temperature or alloy composition, is changed. 
Our numerical calculations can quantitatively explain the previously reported experimental results.
\end{abstract}

\maketitle

\narrowtext

\section{Introduction}
The intersubband absorption linewidth in semiconductor quantum wells (QWs) closely relates to fundamental problems in the physics of optical transition, such as relaxation \cite{Ando:1985}, many-body effects \cite{Nikonov:1997,Warburton:1998}, and disorder \cite{Riemann:2002,Luin:2001}. Furthermore, it is a key factor in improving the performance of quantum cascade lasers \cite{Faist:1994} and QW infrared photodetectors \cite{Levine:1993}. 

To investigate the effects of various scattering processes, intersubband absorption linewidths have been measured for various temperatures \cite{Allmen:1988}, well widths \cite{Campman:1996}, alloy compositions \cite{Campman:1996}, and doping positions \cite{Dupont:1992} in GaAs and other QWs. These results show that absorption linewidth has a weak dependence on temperature and alloy composition and apparently has little correlation with mobility. 
Its strong well-width dependence suggests that the main contribution is from interface roughness scattering. 

In a previous paper \cite{Unuma:2001}, we discussed the effect of interface roughness scattering on linewidth by comparing calculations based on a microscopic theory by Ando \cite{Ando:1985} and experimental data for modulation-doped GaAs/AlAs QWs with a well width of $80 \,\mathrm{\AA}$. The results made it clear that linewidth is much more sensitive to interface roughness scattering than transport mobility is, because the contribution from the intrasubband scattering in the first excited subband is much larger than that in the ground subband \cite{Unuma:2001}. Even in wide GaAs QWs, where interface roughness scattering should be less effective, recent reports \cite{Williams:2001,Ullrich:2001} showed that interface roughness scattering has a larger effect on linewidth than either electron-electron scattering or bulk impurity scattering.

In the present paper, we apply our theoretical method \cite{Unuma:2001} to scattering by LO phonons, LA phonons, alloy disorder, and ionized impurities as well as interface roughness scattering, in order to compare their respective contributions to intersubband absorption linewidth and transport mobility. Numerical calculations for GaAs QWs confirm that the very high sensitivity of linewidth to interface roughness scattering is the key to quantitatively explaining the previously reported experimental results for linewidth in comparison with mobility. 

The method presented here follows Ando's theory \cite{Ando:1985}, in which the intrasubband and intersubband energy-dependent single-particle \cite{note1} relaxation rates for various scattering mechanisms are first calculated and then included in a formula for the two-dimensional (2D) dynamical conductivity ${\mathrm{Re}}\,\sigma_{zz}(\omega)$ to give the absorption lineshape for photon frequency $\omega$. 
This method is similar to a familiar method of calculating transport mobility \cite{Ando:1982,Hirakawa:1986}. 
It is important to note that intersubband optical absorption is the collective excitation among a confined electron gas. 
However, our present calculation of single-particle relaxation rates and lineshape is very important and useful, because the absorption lineshape ${\mathrm{Re}}\,\tilde{\sigma}_{zz}(\omega)$ of collective excitation is given by the single-particle dynamical conductivity $\sigma_{zz}(\omega)$ via \cite{Ando:1977}
\begin{eqnarray}
\tilde{\sigma}_{zz}(\omega) 
= \frac{\sigma_{zz}(\omega)}
{1 + \displaystyle 
  \frac{i}{\epsilon_0 \kappa_0 \omega d_{\mathrm{eff}}} \sigma_{zz}(\omega) } 
\end{eqnarray}
in the crudest approximation. 
Here, $\epsilon_0$ is the vacuum permittivity, $\kappa_0$ is the static dielectric constant of the 2D material, and $d_{\mathrm{eff}}$ is the effective thickness of the 2D electron gas \cite{Ando:1982}.

The collective excitation effects, or many-body interaction effects, on intersubband absorption linewidth have been issues of recent interest in both theoretical and experimental studies. In the limit of small band-nonparabolicity and constant single-particle relaxation rates, Nikonov \textit{et al.} theoretically showed that many-body effects only cause blue-shifts in absorption spectra (the depolarization shift) and that the linewidth is solely determined by the single-particle relaxation rate \cite{Nikonov:1997}. 

In largely nonparabolic systems, the variation in energy separation between the ground and first excited subbands produces additional width in the single-particle excitation lineshape ${\mathrm{Re}}\,\sigma_{zz}(\omega)$. 
However, many-body effects lead to redistribution of oscillator strength and collective excitation that has a sharp resonance. 
As a result, the linewidth of the collective excitation spectrum ${\mathrm{Re}}\,\tilde{\sigma}_{zz}(\omega)$ is significantly different from that of the single-particle excitation spectrum ${\mathrm{Re}}\,\sigma_{zz}(\omega)$ \cite{Nikonov:1997,Warburton:1996,Zaluzny:1991}. 
Furthermore, nonparabolicity causes difficulties in calculating single-particle relaxation rates and ${\mathrm{Re}}\,\sigma_{zz}(\omega)$ by Ando's formalism \cite{Ando:1978}. 
Experiments to elucidate these effects were performed by Warburton \textit{et al.} on InAs/AlSb QWs \cite{Warburton:1998}. 

In the more popular systems of GaAs/AlGaAs and InGaAs/InAlAs QWs, however, our present calculation, which assumes small nonparabolicity, is applicable. 
The purposes of this paper are (1) to calculate intrasubband and intersubband single-particle relaxation rates for relevant scattering mechanisms as functions of in-plane kinetic energy assuming small nonparabolicity and (2) to quantitatively explain previously reported experimental data on linewidth and mobility in GaAs-based QWs, which appeared to have little correlation.

In the next section, we summarize how linewidth and mobility are related to single-particle relaxation rates, and calculate the single-particle relaxation rates for various scattering mechanisms as functions of the kinetic energy $E$. 
It is shown that linewidth and mobility have very different sensitivities to the same scattering mechanism. 
In Section 3, previously reported experimental data for various temperatures, well widths, alloy compositions, and doping positions are quantitatively explained by numerical calculations.

\section{Formulation of the problem}
\subsection{General theory of intersubband absorption lineshape and transport mobility in quantum wells}

A general theory of intersubband absorption linewidth due to elastic scatterers in 2D systems was formulated by Ando \cite{Ando:1985,Ando:1978}. 
According to Ando's theory, the absorption lineshape of single-particle excitation between the two lowest subbands can be expressed as
\begin{eqnarray}
{\mathrm{Re}} \,\sigma_{zz}(\omega)
= \frac{e^2 f_{10}}{2m^*} \int \frac{m^*}{\pi \hbar^2} dE f(E) 
\frac{\hbar \Gamma_{\mathrm{op}}(E)}{(\hbar\omega - E_{10})^2 
 + \Gamma_{\mathrm{op}}(E)^2},
\label{conductivity}
\end{eqnarray}
when all electrons are initially in the ground subband. Here, 
\begin{eqnarray}
\Gamma_{\mathrm{op}}(E) &=& \frac{1}{2} \left[ 
  \Gamma_{\mathrm{intra}}(E) + \Gamma_{\mathrm{inter}}(E) \right], \\
\Gamma_{\mathrm{intra}}(E) &=& 2\pi \displaystyle\sum_{{\mathbf{k}}^{\prime}} \left. \left<\,
  |(0{\mathbf{k}}^{\prime}|H_1|0{\mathbf{k}}) - (1{\mathbf{k}}^{\prime}|H_1|1{\mathbf{k}})
  |^2 \,\right> 
  \delta\left(\varepsilon({\mathbf{k}}) - \varepsilon({\mathbf{k}}^{\prime})\right)
  \,\right|_{E=\varepsilon({\mathbf{k}})},   \label{intra} \\
\Gamma_{\mathrm{inter}}(E) &=& 2\pi \displaystyle\sum_{{\mathbf{k}}^{\prime}} \left. \left<\,
  |(0{\mathbf{k}}^{\prime}|H_1|1{\mathbf{k}}) |^2 \,\right> 
  \delta \left(\varepsilon({\mathbf{k}}) - \varepsilon({\mathbf{k}}^{\prime}) + E_{10}
  \right)  \,\right|_{E=\varepsilon({\mathbf{k}})},  \label{inter}
\end{eqnarray}
$e$ is the elementary charge, $\hbar$ is the reduced Planck constant, $m^*$ is the electron effective mass, $f_{10}$ is the oscillator strength, $E_{10}$ ($= E_1 - E_0$) is the intersubband energy separation, $f(E)$ is the Fermi distribution function at temperature $T$, $|n{\mathbf{k}})$ is the state vector of the electron with subband index $n$ and wave vector ${\mathbf{k}}$, $E_n$ is the quantization energy, $\varepsilon({\mathbf{k}})= \hbar^2 k^2/2m^*$,  $H_1$ is the scattering potential, and $\left< \cdots \right>$ denotes the average over distribution of scatterers. 
This theory assumes a parabolic conduction band, or a constant effective mass for different subbands; a modification for slightly nonparabolic systems like GaAs QWs will be described in a later paragraph. 
In this paper we denote the full width at half maximum of the spectrum given by Eq. (\ref{conductivity}) as $2\Gamma_{\mathrm{op}}$. 

Note, on the other hand, that the transport relaxation time $\tau_{\mathrm{tr}}(E)$, or the transport relaxation rate $2\Gamma_{\mathrm{tr}}(E) = 2\hbar/\tau_{\mathrm{tr}}(E)$ can be expressed as \cite{Ando:1982}
\begin{eqnarray}
\frac{2 \hbar}{\tau_{\mathrm{tr}}(E)} 
= 4\pi \displaystyle\sum_{{\mathbf{k}}^{\prime}} \left. \left<\,
  |(0{\mathbf{k}}^{\prime}|H_1|0{\mathbf{k}}) |^2 \,\right> 
  \delta \left(\varepsilon({\mathbf{k}}) - \varepsilon({\mathbf{k}}^{\prime}) \right)
  (1 - \cos\theta)  \,\right|_{E=\varepsilon({\mathbf{k}})},  \label{transport}
\end{eqnarray}
where $\theta$ is the angle between ${\mathbf{k}}$ and ${\mathbf{k}}^\prime$. 
The mobility is given by $\mu = e \tau_{\mathrm{tr}}/m^*$ with an average relaxation time of \cite{Ando:1982,Hirakawa:1986}
\begin{eqnarray}
\tau_{\mathrm{tr}}
= \int dE \tau_{\mathrm{tr}}(E) E \frac{\partial f(E)}{\partial E} \left/
  \int dE E \frac{\partial f(E)}{\partial E}. \right.
\end{eqnarray}
To enable quantitative comparison between the linewidth $2\Gamma_{\mathrm{op}}$ and mobility $\mu$, we define the transport energy broadening as $2\Gamma_{\mathrm{tr}} = 2\hbar/\tau_{\mathrm{tr}} = 2\hbar e/m^* \mu$ \cite{Unuma:2001}. 
In particular, low-temperature transport broadening is given by $2\Gamma_{\mathrm{tr}} = 2\Gamma_{\mathrm{tr}}(E_F) = 2\hbar/\tau_{\mathrm{tr}}(E_F)$, where $E_F$ is the 2D Fermi energy.

There are two relevant many-body effects: static and dynamic screening. The former screens the potentials of elastic scatterers while the latter induces collective charge-density excitation because of the incident optoelectric field.

The static screening effect can be included by replacing the scattering matrix element $(m{\mathbf{k}}^{\prime}|H_1|n{\mathbf{k}})$ with \cite{Ando:1985}
\begin{eqnarray}
(m{\mathbf{k}}^{\prime}|H_1|n{\mathbf{k}}) + (0{\mathbf{k}}^{\prime}|H_1|0{\mathbf{k}})
\left[ \frac{1}{\epsilon(q,T)} - 1 \right] 
  \frac{F_{(00)(mn)}(q)}{F_{(00)(00)}(q)}.
\end{eqnarray}
Here, ${\mathbf{q}} = {\mathbf{k}} - {\mathbf{k}}^{\prime}$, $\epsilon(q,T)$ is the static dielectric function \cite{Ando:1982,Hirakawa:1986}, and $F_{(kl)(mn)}(q)$ is a form factor defined by \cite{Ando:1985}
\begin{eqnarray}
F_{(kl)(mn)}(q) = \int dz \int dz^\prime \,
  \zeta_k(z) \zeta_l(z) \zeta_m(z^\prime) \zeta_n(z^\prime) 
  \, e^{-q|z - z^\prime|}.
\end{eqnarray}
The $z$ axis is set along the growth direction of samples, and $\zeta_n(z)$ is the wave function for the $n$-th subband electron motion in the $z$ direction, which is chosen to be real. 
The screening correction only results in dividing $(0{{\mathbf{k}}^\prime}|H_1|0{\mathbf{k}})$ in Eq. (\ref{intra}) by the factor
\begin{eqnarray}
S(q,T) = \left[ \frac{1}{\epsilon(q,T)} 
  - \left( \frac{1}{\epsilon(q,T)} - 1 \right)
  \frac{F_{(00)(11)}(q)}{F_{(00)(00)}(q)} \right]^{-1},
\end{eqnarray}
and $(0{{\mathbf{k}}^\prime}|H_1|0{\mathbf{k}})$ in Eq. (\ref{transport}) by $\epsilon(q,T)$. 
In this paper we only treat symmetrical QWs, so there is no screening factor in Eq. (\ref{inter}). 
$\epsilon(q,T)$ significantly increases mobility, while $S(q,T)$ hardly affects absorption linewidth because $S(q,T) \sim 1$.

The dynamic screening effect is counted as a depolarization field, and the absorption lineshape ${\mathrm{Re}}\,\tilde{\sigma}_{zz}(\omega)$ of the induced collective charge-density excitation is given by 
\begin{eqnarray}
\tilde{\sigma}_{zz}(\omega) 
= \frac{\sigma_{zz}(\omega)}{\epsilon_{zz}(\omega)}
\end{eqnarray}
with the dynamical dielectric function of
\begin{eqnarray}
\epsilon_{zz}(\omega) 
= 1 + \frac{i}{\epsilon_0 \kappa_0 \omega d_{\mathrm{eff}}}
       \sigma_{zz}(\omega).
\end{eqnarray}
The resonance energy $\tilde{E}_{10}$ of ${\mathrm{Re}}\,\tilde{\sigma}_{zz}(\omega)$ is blue-shifted from the original resonance energy $E_{10}$ of ${\mathrm{Re}}\,\sigma_{zz}(\omega)$, and 
\begin{eqnarray}
{\tilde{E}}_{10} = \sqrt{E_{10}{}^2 + (\hbar\omega_p)^2}
\end{eqnarray}
with the plasma frequency of
\begin{eqnarray}
\omega_p = \sqrt{\frac{f_{10} N_S e^2}
{\epsilon_0 \kappa_0 m^* d_{\mathrm{eff}}}}.
\end{eqnarray}
The blue-shift ${\tilde{E}}_{10} - E_{10} \approx (\hbar \omega_p)^2/(2E_{10})$ is called the depolarization shift. 
The linewidth of ${\mathrm{Re}}\,\tilde{\sigma}_{zz}(\omega)$ is the same as that of ${\mathrm{Re}}\,\sigma_{zz}(\omega)$ if $2\Gamma_{\mathrm{op}}(E)$ is independent of energy \cite{Nikonov:1997}, though they are different in general. 
When the depolarization shift is sufficiently small, or
\begin{eqnarray}
{\tilde{E}}_{10} - E_{10} < 2\Gamma_{\mathrm{op}}(0),
\label{small depolarization}
\end{eqnarray}
$\tilde{\sigma}_{zz}(\omega)$ is approximately equal to $\sigma_{zz}(\omega)$. 

Although Eqs. (\ref{conductivity})-(\ref{inter}) were derived assuming parabolic bands, we may apply them to slightly nonparabolic systems in which the additional width due to nonparabolicity is small compared with the width due to scattering mechanisms. 
The condition is expressed as $(1 - m^*_0/m^*_1)E_F < 2\Gamma_{\mathrm{op}}(0)$ at low temperatures, where $m^*_n$ is the electron effective mass in the $n$-th subband. 
In this case, we can use the present theory by replacing $E_{10}$ in Eq. (\ref{conductivity}) with $E_{10}(0) - (1 - m^*_0/m^*_1)E$ \cite{Zaluzny:1991,Ando:1978-2}, which has a much larger influence on absorption linewidth than other corrections. Here, $E_{10}(0)$ represents the intersubband energy separation at ${\mathbf{k}} = 0$. 
For consistency, respective $\delta$-functions appearing with the squares of scattering matrix elements 
$|(0{{\mathbf{k}}^\prime}|H_1|0{\mathbf{k}})|^2$, 
$|(1{{\mathbf{k}}^\prime}|H_1|1{\mathbf{k}})|^2$, 
$|(0{{\mathbf{k}}^\prime}|H_1|0{\mathbf{k}})(1{{\mathbf{k}}^\prime}|H_1|1{\mathbf{k}})|$, and
$|(0{{\mathbf{k}}^\prime}|H_1|1{\mathbf{k}})|^2$
in Eqs. (\ref{intra}) and (\ref{inter}) should be replaced by 
$\delta(\varepsilon_{0}({\mathbf{k}}) - \varepsilon_{0}({\mathbf{k}}^\prime))$, 
$\delta(\varepsilon_{1}({\mathbf{k}}) - \varepsilon_{1}({\mathbf{k}}^\prime))$, 
$\frac{1}{2} 
\left[ \delta(\varepsilon_{0}({\mathbf{k}})- \varepsilon_{0}({\mathbf{k}}^\prime))
+ \delta(\varepsilon_{1}({\mathbf{k}}) - \varepsilon_{1}({\mathbf{k}}^\prime)) \right]$, 
and $\delta(\varepsilon_{1}({\mathbf{k}}) - \varepsilon_{0}({\mathbf{k}}^\prime))$, 
where $\varepsilon_{n}({\mathbf{k}}) = E_n + \hbar^2 k^2/2m^*_n$. 
Values of $m^*_n$ obtained from the Kane model are used in our numerical calculations. 

In most cases of GaAs QWs examined later in this paper, the depolarization shift ${\tilde{E}}_{10} - E_{10}$ and the nonparabolicity effect $(1 - m^*_0/m^*_1)E_F$ are small compared with $2\Gamma_{\mathrm{op}}(0)$, so the absorption linewidth is estimated directly from ${\mathrm{Re}}\,\sigma_{zz}(\omega)$ in Eq. (\ref{conductivity}).

\subsection{Scattering mechanisms}
In this section, we calculate and compare $2\Gamma_{\mathrm{op}}(E)$ and $2\Gamma_{\mathrm{tr}}(E)$ due to scattering by interface roughness (IFR), LO phonons, LA phonons, alloy disorder (AD), and ionized impurities (ION). Furthermore, numerical calculations of each individual scattering mechanism are performed for modulation-doped GaAs (or InGaAs)/AlAs QWs. 
In actual samples, several scattering mechanisms coexist; the total scattering rate can be obtained as the sum of their rates. Namely, 
\begin{eqnarray}
\Gamma_{\mathrm{op}}(E) 
&=& \Gamma_{\mathrm{op}}^{\mathrm{(IFR)}}(E)
+ \Gamma_{\mathrm{op}}^{\mathrm{(LO)}}(E)
+ \Gamma_{\mathrm{op}}^{\mathrm{(LA)}}(E)
+ \Gamma_{\mathrm{op}}^{\mathrm{(AD)}}(E)
+ \Gamma_{\mathrm{op}}^{\mathrm{(ION)}}(E)
+ \cdots, \\
\Gamma_{\mathrm{tr}}(E) 
&=& \Gamma_{\mathrm{tr}}^{\mathrm{(IFR)}}(E)
+ \Gamma_{\mathrm{tr}}^{\mathrm{(LO)}}(E)
+ \Gamma_{\mathrm{tr}}^{\mathrm{(LA)}}(E)
+ \Gamma_{\mathrm{tr}}^{\mathrm{(AD)}}(E)
+ \Gamma_{\mathrm{tr}}^{\mathrm{(ION)}}(E)
+ \cdots .
\end{eqnarray}

For simplicity, we perform numerical calculations in single QWs with a finite barrier height of $V_0$ where band bending due to doping is neglected. The origin of the $z$ axis is set at the center of the QWs. Material constants of GaAs used in calculations are shown in Table \ref{GaAs material constant}.

\subsubsection{Interface roughness scattering}
In GaAs/AlGaAs QWs, dominant monolayer (ML) fluctuations are formed at the GaAs-on-AlGaAs interface (AlGaAs surface covered by GaAs). 
We assume that the roughness height $\Delta({\mathbf{r}})$ at the in-plane position ${\mathbf{r}}=(x,y)$ has a correlation function \cite{Ando:1982,Sakaki:1987}:
\begin{eqnarray}
  \left< \Delta({\mathbf{r}}) \Delta({\mathbf{r}}^{\prime}) \right> 
  = \Delta^2 \exp \left( -\frac{|{\mathbf{r}}-{\mathbf{r}}^{\prime}|^2}{\Lambda^2} \right),  \end{eqnarray}
where $\Delta$ is the mean height of roughness and $\Lambda$ is the correlation length. 
The scattering matrix element is given by
\begin{eqnarray}
(m{\mathbf{k}}^{\prime}|H_1|n{\mathbf{k}}) = \int d^2r
   F_{mn}\,\Delta({\mathbf{r}}) \,e^{i{\mathbf{q}}\cdot{\mathbf{r}}} \label{ifr-matrix}
\end{eqnarray}
with 
\begin{eqnarray}
F_{mn} = V_0 \,\zeta_m(-L/2) \,\zeta_n(-L/2),  \label{effective field}
\end{eqnarray}
where $L$ is the well width and $\zeta_n(-L/2)$ is the wave function at the GaAs-on-AlGaAs interface. Because interface roughness is equivalent to local fluctuations in well width, $F_{mn}$ in Eq. (\ref{effective field}) can also be expressed as 
\begin{eqnarray}
F_{mn} = \sqrt{(\partial E_m/\partial L)(\partial E_n/\partial L)}.
\label{effective field 2}
\end{eqnarray}
In the case of the infinite-barrier approximation, Eq. (\ref{effective field}) can be expressed in an alternative form as \cite{Ando:1982,Ando:1976}
\begin{eqnarray}
F_{mn} = \left. \frac{\hbar^2}{2m^*} 
\frac{d \zeta_m(z)}{dz} \frac{d \zeta_n(z)}{dz} \right|_{z = -L/2},
\end{eqnarray}
which is found to be proportional to $L^{-3}$.

Substituting Eq. (\ref{ifr-matrix}) into Eqs. (\ref{intra}) and (\ref{inter}), we get
\begin{eqnarray}
\Gamma_{\mathrm{intra}}^{\mathrm{(IFR)}}(E) &=& 
\frac{m^* \Delta^2 \Lambda^2}{\hbar^2} \int_0^\pi d\theta
  \left[ \frac{F_{00}}{S(q,T)} - F_{11} \right]^2
  e^{-q^2 \Lambda^2/4},  \label{ifr-intra} \\
\Gamma_{\mathrm{inter}}^{\mathrm{(IFR)}}(E) &=&
\frac{m^* \Delta^2 \Lambda^2}{\hbar^2} F_{01}{}^2 \int_0^\pi d\theta\,
  e^{-\tilde{q}^2 \Lambda^2/4},  \label{ifr-inter}
\end{eqnarray}
where the absolute values of the 2D scattering vectors $q$ and $\tilde{q}$ are  given by \cite{Unuma:2001}
\begin{eqnarray}
q^2 &=& 2k^2 (1 - \cos\theta), \\
\tilde{q}^2 &=& 2k^2 + \frac{2m^* E_{10}}{\hbar^2} 
  - 2k\sqrt{k^2 + \frac{2m^* E_{10}}{\hbar^2}} \cos\theta.
\end{eqnarray}

On the other hand, we can express the transport relaxation time $\tau_{\mathrm{tr}}^{\mathrm{(IFR)}}(E)$ \cite{Ando:1982-2}, or the transport relaxation rate $2\Gamma_{\mathrm{tr}}^{\mathrm{(IFR)}}(E) = 2\hbar/\tau_{\mathrm{tr}}^{\mathrm{(IFR)}}(E)$ as
\begin{eqnarray}
2\Gamma_{\mathrm{tr}}^{\mathrm{(IFR)}}(E) =
\frac{2m^* \Delta^2 \Lambda^2}{\hbar^2} F_{00}{}^2
  \int_0^\pi \! d\theta\,
  \frac{1 - \cos\theta}{\epsilon(q,T)^2} 
  \,\, e^{-q^2 \Lambda^2/4},
\label{ifr-tr}
\end{eqnarray}
which is similar to Eqs. (\ref{ifr-intra}) and (\ref{ifr-inter}).

It is useful here to comment on the similarities and differences in the equations for $\Gamma_{\mathrm{intra}}(E)$, $\Gamma_{\mathrm{inter}}(E)$, and $2\Gamma_{\mathrm{tr}}(E)$. 
First, all three are proportional to $\Delta^2$, and also to $\Lambda^2$ for small $\Lambda$. 
Second, $\Gamma_{\mathrm{inter}}(E)$ is much smaller than $\Gamma_{\mathrm{intra}}(E)$, because $\tilde{q}$ is smaller than $q$. 
Third, $(1 - \cos \theta)/\epsilon(q,T)^2$ appearing in $2\Gamma_{\mathrm{tr}}(E)$ shows that the forward scattering ($\theta \sim 0$) does not contribute to transport broadening, and that the screening effect reduces the scattering rates. 
Finally, and most importantly, they include different factors $[F_{00}/S(q,T) - F_{11}]^2$, $F_{01}{}^2$, and $2F_{00}{}^2$. 
$S(q,T)$ can be neglected because $S(q,T) \sim 1$. 
As is shown below, $F_{11}$ is much larger than $F_{00}$, because $E_1$ is more sensitive to $L$ than $E_0$. (In the infinite-barrier approximation, $F_{11}$ is four times larger than $F_{00}$.) 
As a result, $\Gamma_{\mathrm{intra}}(E)$ is much larger than $2\Gamma_{\mathrm{tr}}(E)$.

\begin{figure}[bt]
\begin{center}
\includegraphics[width=.4\textwidth]{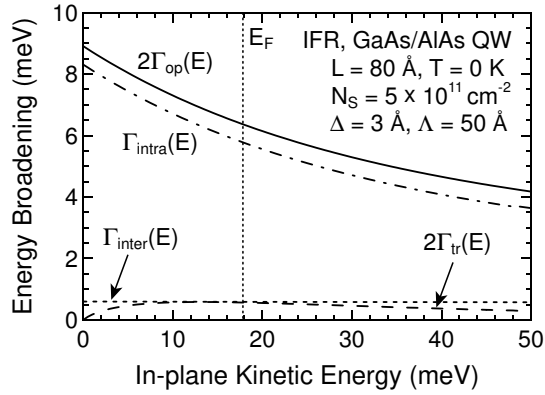}
\caption{$2\Gamma_{\mathrm{op}}(E)$, $\Gamma_{\mathrm{intra}}(E)$, $\Gamma_{\mathrm{inter}}(E)$, and $2\Gamma_{\mathrm{tr}}(E)$ due to interface roughness (IFR) scattering, plotted as functions of the in-plane kinetic energy $E$.}
\label{ifr-energy}
\end{center}
\end{figure}

Figure \ref{ifr-energy} shows $2\Gamma_{\mathrm{op}}(E)$, $\Gamma_{\mathrm{intra}}(E)$, $\Gamma_{\mathrm{inter}}(E)$, and $2\Gamma_{\mathrm{tr}}(E)$ in a modulation-doped GaAs/AlAs QW with $L = 80 \,\mathrm{\AA}$, $\Delta = 3 \,\mathrm{\AA}$, and $\Lambda = 50 \,\mathrm{\AA}$.
These values of $\Delta$ and $\Lambda$ are typical for the GaAs-on-AlAs interface \cite{Unuma:2001,Sakaki:1987,Tanaka:1987}. 
Temperature was set at $T = 0\,\mathrm{K}$, and sheet electron concentration was chosen to be $N_S = 5 \times 10^{11}\,\mathrm{cm}^{-2}$, which gives Fermi energy of $E_F = 17.8\,\mathrm{meV}$. 
The same values of $L$, $N_S$, $\Delta$, and $\Lambda$ are also used for calculations of other scattering mechanisms in this section. 

In Fig. \ref{ifr-energy}, $\Gamma_{\mathrm{intra}}(E)$ decreases as $E$ increases, and it has a maximum value of $8.3 \,\mathrm{meV}$ at $E=0\,\mathrm{meV}$. $\Gamma_{\mathrm{inter}}(E)$ is almost constant with respect to $E$, and its value of $0.6\,\mathrm{meV}$ is much smaller than that of $\Gamma_{\mathrm{intra}}(E)$. 
The values of $2\Gamma_{\mathrm{tr}}(E)$ are $0\,\mathrm{meV}$ at $E=0\,\mathrm{meV}$ owing to the screening effect, and $0.6\,\mathrm{meV}$ at $E = E_F$, which determines the low-temperature transport broadening. 
As a result, $2\Gamma_{\mathrm{op}}(E)$, the sum of $\Gamma_{\mathrm{intra}}(E)$ and $\Gamma_{\mathrm{inter}}(E)$, is found to be much larger than $2\Gamma_{\mathrm{tr}}(E)$.

\begin{figure}[bt]
\begin{center}
\includegraphics[width=.4\textwidth]{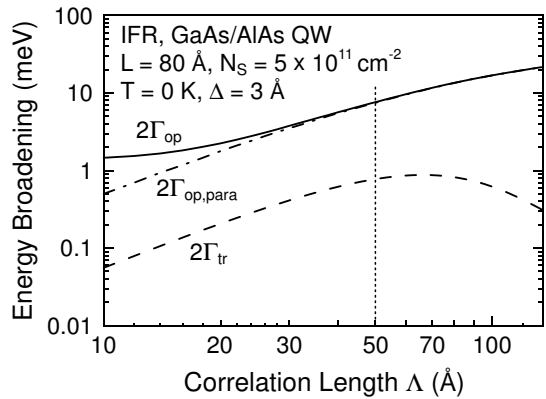}
\caption{Intersubband absorption linewidth $2\Gamma_{\mathrm{op}}$ and the transport energy broadening $2\Gamma_{\mathrm{tr}}$ due to interface roughness (IFR) scattering, plotted as functions of the correlation length $\Lambda$. The $2\Gamma_{\mathrm{op,para}}$ represents the absorption linewidth without the additional width due to band-nonparabolicity.}
\label{ifr-lambda}
\end{center}
\end{figure}

Figure \ref{ifr-lambda} shows $2\Gamma_{\mathrm{op}}$ and $2\Gamma_{\mathrm{tr}}$ as functions of correlation length $\Lambda$; they are calculated respectively from Eqs. (\ref{conductivity}) and (\ref{transport}) in a modulation-doped GaAs/AlAs QW with $L = 80 \,\mathrm{\AA}$, $N_S = 5 \times 10^{11}\,\mathrm{cm}^{-2}$, $T = 0\,\mathrm{K}$, and $\Delta = 3 \,\mathrm{\AA}$. $2\Gamma_{\mathrm{op,para}}$ represents the linewidth calculated without changing $E_{10}$ into $E_{10}(0) - (1 - m^*_0/m^*_1)E$ in Eq. (\ref{conductivity}).

In Fig. \ref{ifr-lambda}, $2\Gamma_{\mathrm{op,para}}$ and $2\Gamma_{\mathrm{tr}}$ are both proportional to $\Lambda^2$ for small $\Lambda$ with the difference in absolute values being about one order of magnitude. 
With nonparabolicity, $2\Gamma_{\mathrm{op}}$ has a lower limit of $(1 - m_0^*/m_1^*)E_F = 1.35\,\mathrm{meV}$ in addition to $2\Gamma_{\mathrm{op,para}}$. 
For large $\Lambda$, the insensitivity of $2\Gamma_{\mathrm{tr}}$ to forward scattering causes its value to be smaller. This shows that the correlation length of $\Lambda \sim 1/k_F$ contributes most to $2\Gamma_{\mathrm{tr}}$, where $k_F$ is the Fermi wavenumber. 
In principle, values of the roughness parameters $\Delta$ and $\Lambda$ can be uniquely determined if linewidth and mobility are both measured at low temperatures.

As shown above, the main characteristic of interface roughness scattering is its order-of-magnitude different contributions to linewidth and transport broadening (and hence to mobility). This is the key point for understanding the apparent lack of correlation between them.

\subsubsection{LO phonon scattering}
In considering phonon scattering processes, it should be noted that phonons have approximately three-dimensional (3D) properties, since they are hardly confined to QWs. 
The $z$-component momentum conservation in 3D systems requires the scattering matrix element of 2D electrons to be calculated from \cite{Price:1981}
\begin{eqnarray}
|M_{\mathrm{2D}}|^2 = \sum_{q_z} |M_{\mathrm{3D}} I(q_z)|^2.
\end{eqnarray}
Here, $M_{\mathrm{2D}}$ and $M_{\mathrm{3D}}$ are the 2D and 3D scattering matrix elements, respectively, and $I(q_z)$ is given by 
\begin{eqnarray}
I(q_z) = I_{mn}(q_z) = \int dz \,\zeta_m(z) \zeta_n(z) \,e^{iq_z z}.
\end{eqnarray}
Since the method of calculating the 3D scattering matrix element is well established, the 2D scattering matrix element can be easily obtained. 

In polar optical phonon scattering, or simply LO phonon scattering, the 3D scattering matrix element is given by \cite{Price:1981}
\begin{eqnarray}
\left< |M_{\mathrm{3D}}|^2 \right> 
= \frac{e^2 \hbar \omega_{\mathrm{LO}} \left( N_{\mathrm{LO}} \!+\! 1 \right)}
  {2\epsilon_0 Q^2}
  \left( \frac{1}{\kappa_\infty} - \frac{1}{\kappa_0} \right)
\end{eqnarray}
for phonon emission processes, and by
\begin{eqnarray}
\left< |M_{\mathrm{3D}}|^2 \right> 
= \frac{e^2 \hbar \omega_{\mathrm{LO}} N_{\mathrm{LO}}}{2\epsilon_0 Q^2}
\left( \frac{1}{\kappa_\infty} - \frac{1}{\kappa_0} \right)
\end{eqnarray}
for phonon absorption processes.
Here, $Q$ is the absolute value of the 3D scattering vector, $\kappa_\infty$ is the optical dielectric constant, 
$\omega_{\mathrm{LO}}$ is the LO phonon frequency, and $N_{\mathrm{LO}}$ is the LO phonon occupation given by 
\begin{eqnarray}
N_{\mathrm{LO}} 
= \frac{1}{ e^{\hbar \omega_{\mathrm{LO}}/k_{\mathrm{B}} T} - 1 }.
\end{eqnarray}
Since LO phonon scattering is an inelastic process, Eqs. (\ref{intra}) and (\ref{inter}) are not applicable in their original forms. 
However, by modifying the $\delta$-functions in Eqs. (\ref{intra}) and (\ref{inter}), 
\begin{eqnarray}
\delta \left( \varepsilon({\mathbf{k}}) - \varepsilon({{\mathbf{k}}^\prime}) \right)
&\longrightarrow&
\delta \left( \varepsilon({\mathbf{k}}) - \varepsilon({{\mathbf{k}}^\prime})
  \pm \hbar\omega_{\mathrm{LO}} \right), \\
\delta \left( \varepsilon({\mathbf{k}}) - \varepsilon({{\mathbf{k}}^\prime})
  + E_{10} \right)
&\longrightarrow&
\delta \left( \varepsilon({\mathbf{k}}) - \varepsilon({{\mathbf{k}}^\prime})
  + E_{10} \pm \hbar\omega_{\mathrm{LO}} \right), 
\end{eqnarray}
such that total energy is conserved, we can estimate width of the zero-phonon band in an approximation that neglects phonon sidebands. Here, $\pm$ indicates phonon absorption ($+$) and emission ($-$). Thus, we have 
\begin{eqnarray}
\Gamma_{\mathrm{intra}}^{\mathrm{(LO)}}(E) &=&
  \frac{m^* e^2 \omega_{\mathrm{LO}}}{4\pi \epsilon_0 \hbar}
  \left( \frac{1}{\kappa_\infty} - \frac{1}{\kappa_0} \right)
  \int_0^\pi d\theta \nonumber \\ && \times \left[ 
  \Theta(E - \hbar\omega_{\mathrm{LO}})
  \frac{\left< N_{\mathrm{LO}} \!+\! 1 \right>}{q_{\mathrm{e}}}
  \left\{  F_{(00)(00)}(q_{\mathrm{e}})
  - 2F_{(00)(11)}(q_{\mathrm{e}})
  + F_{(11)(11)}(q_{\mathrm{e}})  \right\} \right. 
  \nonumber \\ && \quad \!+ \left.
  \frac{\left< N_{\mathrm{LO}} \right>}{q_{\mathrm{a}}}
  \left\{  F_{(00)(00)}(q_{\mathrm{a}})
  - 2F_{(00)(11)}(q_{\mathrm{a}})
  + F_{(11)(11)}(q_{\mathrm{a}})  \right\} \right],
  \label{LO-intra}\\
\Gamma_{\mathrm{inter}}^{\mathrm{(LO)}}(E) &=&
  \frac{m^* e^2 \omega_{\mathrm{LO}}}{4\pi \epsilon_0 \hbar}
  \left( \frac{1}{\kappa_\infty} - \frac{1}{\kappa_0} \right)
  \int_0^\pi d\theta \nonumber \\ && \times \left[ 
  \Theta(E + E_{10} - \hbar\omega_{\mathrm{LO}})
  \frac{\left< N_{\mathrm{LO}} \!+\! 1 \right>}{\tilde{q}_{\mathrm{e}}}
  F_{(01)(10)}(\tilde{q}_{\mathrm{e}})
  + \frac{\left< N_{\mathrm{LO}} \right>}{\tilde{q}_{\mathrm{a}}}
  F_{(01)(10)}(\tilde{q}_{\mathrm{a}})  \right],  \label{LO-inter}
\end{eqnarray}
where $\Theta(E)$ is the Heaviside step function. Absolute values of scattering vectors are given by
\begin{eqnarray}
q_{\mathrm{e}}{}^2 
  &=&  2k^2 - \frac{2m^* \omega_{\mathrm{LO}}}{\hbar} 
  - 2k\sqrt{ k^2 - \frac{2m^* \omega_{\mathrm{LO}}}{\hbar} }\cos\theta, \\
q_{\mathrm{a}}{}^2 
  &=&  2k^2 + \frac{2m^* \omega_{\mathrm{LO}}}{\hbar} 
  - 2k\sqrt{ k^2 + \frac{2m^* \omega_{\mathrm{LO}}}{\hbar} }\cos\theta, \\
\tilde{q}_{\mathrm{e}}{}^2 
  &=&  2k^2 + \frac{2m^* E_{10}}{\hbar^2} 
  - \frac{2m^* \omega_{\mathrm{LO}}}{\hbar} 
  - 2k\sqrt{ k^2 + \frac{2m^* E_{10}}{\hbar^2} 
    - \frac{2m^* \omega_{\mathrm{LO}}}{\hbar} }\cos\theta,  \\
\tilde{q}_{\mathrm{a}}{}^2 &=&  2k^2 + \frac{2m^* E_{10}}{\hbar^2} 
  + \frac{2m^* \omega_{\mathrm{LO}}}{\hbar} 
  - 2k\sqrt{ k^2 + \frac{2m^* E_{10}}{\hbar^2} 
    + \frac{2m^* \omega_{\mathrm{LO}}}{\hbar} }\cos\theta, 
\end{eqnarray}
and the subscripts ``e'' and ``a'' represent emission and absorption of LO phonons, respectively.

On the other hand, the transport relaxation rate can be expressed as
\begin{eqnarray}
2\Gamma_{\mathrm{tr}}^{\mathrm{(LO)}}(E) &=&
  \frac{m^* e^2 \omega_{\mathrm{LO}}}{2\pi \epsilon_0 \hbar}
  \left( \frac{1}{\kappa_\infty} - \frac{1}{\kappa_0} \right)
  \frac{1}{1 - f(E)} \int_0^\pi d\theta
\nonumber \\ && \times \left[ 
  \Theta(E - \hbar\omega_{\mathrm{LO}})
  \left\{ 1 - f(E - \hbar \omega_{\mathrm{LO}}) \right\}
  \frac{\left< N_{\mathrm{LO}} \!+\! 1 \right>}{q_{\mathrm{e}}}
  F_{(00)(00)}(q_{\mathrm{e}}) \right. 
  \nonumber \\ && \quad
  \left. + \left\{ 1 - f(E + \hbar \omega_{\mathrm{LO}}) \right\}
  \frac{\left< N_{\mathrm{LO}} \right>}{q_{\mathrm{a}}}
  F_{(00)(00)}(q_{\mathrm{a}})  \right] 
  \label{LO-tr}
\end{eqnarray}
in the approximation that neglects the ``in-scattering term''\cite{Hirakawa:1986}. Since LO phonon frequency is high, the screening effect can be neglected.

\begin{figure}[bt]
\begin{center}
\includegraphics[width=.4\textwidth]{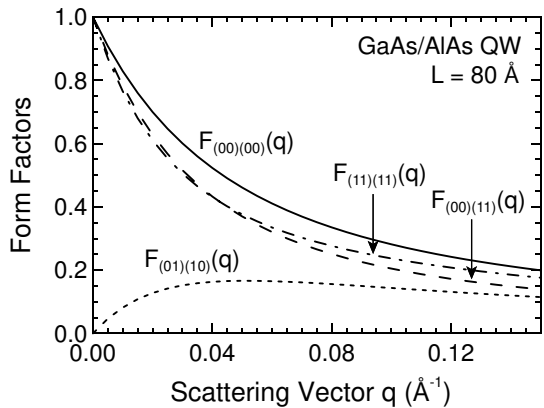}
\caption{Dependence of the form factors $F_{(00)(00)}(q)$, $F_{(00)(11)}(q)$, $F_{(11)(11)}(q)$, and $F_{(01)(10)}(q)$ on the absolute value of the two-dimensional scattering vector $q$.}
\label{formfactors}
\end{center}
\end{figure}

The four form factors $F_{(00)(00)}(q)$, $F_{(00)(11)}(q)$, $F_{(11)(11)}(q)$, and $F_{(01)(10)}(q)$ appearing in Eqs. (\ref{LO-intra}), (\ref{LO-inter}), and (\ref{LO-tr}) are plotted as functions of $q$ in Fig. \ref{formfactors}. 
First, $F_{(00)(00)}(q)$, $F_{(00)(11)}(q)$, and $F_{(11)(11)}(q)$ are very close, which makes $F_{(00)(00)}(q) - 2F_{(00)(11)}(q) + F_{(11)(11)}(q)$ in $\Gamma_{\mathrm{intra}}(E)$ much smaller than $F_{(00)(00)}(q)$ in $2\Gamma_{\mathrm{tr}}(E)$. In other words, the difference in intrasubband scattering matrix elements for the two subbands is small in LO phonon scattering. 
Second, $F_{(01)(10)}(q)$ in $\Gamma_{\mathrm{inter}}(E)$ is much smaller than $F_{(00)(00)}(q)$. 

\begin{figure}[bt]
\begin{center}
\includegraphics[width=.4\textwidth]{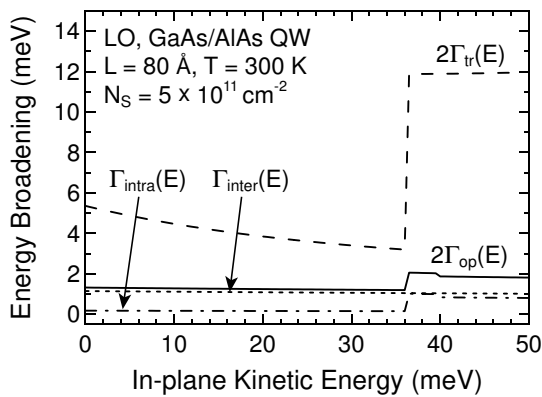}
\caption{$2\Gamma_{\mathrm{op}}(E)$, $\Gamma_{\mathrm{intra}}(E)$, $\Gamma_{\mathrm{inter}}(E)$, and $2\Gamma_{\mathrm{tr}}(E)$ due to LO phonon scattering, plotted as functions of the in-plane kinetic energy $E$. LO phonon energy is $\hbar\omega_{\mathrm{LO}} = 36.5\,\mathrm{meV}$.}
\label{LO-energy}
\end{center}
\end{figure}

Figure \ref{LO-energy} shows $2\Gamma_{\mathrm{op}}(E)$, $\Gamma_{\mathrm{intra}}(E)$, $\Gamma_{\mathrm{inter}}(E)$, and $2\Gamma_{\mathrm{tr}}(E)$ due to LO phonon scattering at $T = 300\,\mathrm{K}$. 
First, $\Gamma_{\mathrm{intra}}(E)$ is much smaller than $2\Gamma_{\mathrm{tr}}(E)$, because the difference in intrasubband scattering matrix elements for the two subband is small in LO phonon scattering as already shown in Fig. \ref{formfactors}. Second, $\Gamma_{\mathrm{inter}}(E)$ is much smaller than $2\Gamma_{\mathrm{tr}}(E)$ owing to the small form factor of $F_{(01)(10)}(q)$ and the large absolute value of scattering vector $\tilde{q}$. 
Third, when the kinetic energy $E$ is larger than the LO phonon energy of $E_{\mathrm{LO}} = 36.5\,\mathrm{meV}$, intrasubband LO phonon emission is allowed, which makes $2\Gamma_{\mathrm{tr}}(E)$ and $\Gamma_{\mathrm{intra}}(E)$ larger. 
As a result, $2\Gamma_{\mathrm{op}}(E)$ is much smaller than $2\Gamma_{\mathrm{tr}}(E)$ at room temperature.

When systems are cooled down to $0\,\mathrm{K}$, only intersubband LO phonon spontaneous emission is allowed in the case of $E_{10} > E_{\mathrm{LO}}$. 
Therefore, $\Gamma_{\mathrm{intra}}(E)$ and $2\Gamma_{\mathrm{tr}}(E)$ vanish, and only $\Gamma_{\mathrm{inter}}(E)$ has a finite value of about $1\,\mathrm{meV}$.

\subsubsection{LA phonon scattering}
Acoustic phonon scattering via deformation potential coupling, or simply LA phonon scattering, is virtually elastic. 
The 3D scattering matrix element is given by \cite{Price:1981}
\begin{eqnarray}
\left< |M_{\mathrm{3D}}|^2 \right> = \frac{k_B TD^2}{2 c_l}
\label{3D LA phonon scattering}
\end{eqnarray}
for both LA phonon emission and absorption processes, where $D$ is the deformation potential constant and $c_l$ is the longitudinal elastic constant. 
Note that Eq. (\ref{3D LA phonon scattering}) is independent of the scattering vector as a result of the linear dispersion relation of LA phonons. 
Therefore, we have
\begin{eqnarray}
\Gamma_{\mathrm{intra}}^{\mathrm{(LA)}}(E) &=& 
\frac{m^* k_B TD^2}{\pi \hbar^2 c_l} \int_0^\pi d\theta \int dz
  \left[ \frac{\zeta_0(z)^2}{S(q,T)} 
  - \zeta_1(z)^2 \right]^2, \\
\Gamma_{\mathrm{inter}}^{\mathrm{(LA)}}(E) &=&
  \frac{m^* k_B TD^2}{\pi \hbar^2 c_l} \int_0^\pi d\theta \int dz
  \left[ \zeta_0(z)\zeta_1(z) \right]^2.
\end{eqnarray}
$\Gamma_{\mathrm{inter}}^{\mathrm{(LA)}}(E)$ is independent of the kinetic energy $E$, and $\Gamma_{\mathrm{intra}}^{\mathrm{(LA)}}(E)$ is also almost independent of it because $S(q,T) \sim 1$. 

On the other hand, the transport relaxation rate is given by \cite{Ando:1982,Price:1981} 
\begin{eqnarray}
2\Gamma_{\mathrm{tr}}^{\mathrm{(LA)}}(E) &=&
  \frac{2m^* k_B D^2 T}{\pi \hbar^2 c_l} \int_0^\pi d\theta\, 
  \frac{1 - \cos\theta}{\epsilon(q,T)^2}
  \int dz \, \zeta_0(z)^4,
\end{eqnarray}
which has an energy dependence due to the screening effect. 

Note here that the $z$-integrals of $\zeta_0(z)^4$, $\zeta_1(z)^4$, and $[\zeta_0(z) \zeta_1(z)]^2$ have comparable values ($3/2L$, $3/2L$, and $1/L$, respectively in the infinite-barrier approximation); thus $\Gamma_{\mathrm{intra}}(E)$ and $\Gamma_{\mathrm{inter}}(E)$ are nearly equal, and $2\Gamma_{\mathrm{op}}(E)$, the sum of them, is comparable with $2\Gamma_{\mathrm{tr}}(E)$.

\begin{figure}[bt]
\begin{center}
\includegraphics[width=.4\textwidth]{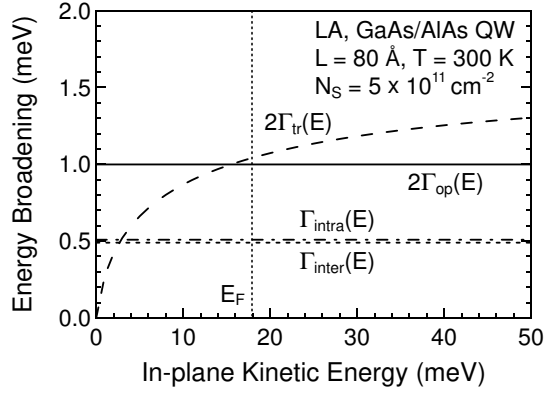}
\caption{$2\Gamma_{\mathrm{op}}(E)$, $\Gamma_{\mathrm{intra}}(E)$, $\Gamma_{\mathrm{inter}}(E)$, and $2\Gamma_{\mathrm{tr}}(E)$ due to LA phonon scattering, plotted as functions of the in-plane kinetic energy $E$.}
\label{LA-energy}
\end{center}
\end{figure}

Figure \ref{LA-energy} shows $2\Gamma_{\mathrm{op}}(E)$, $\Gamma_{\mathrm{intra}}(E)$, $\Gamma_{\mathrm{inter}}(E)$, and $2\Gamma_{\mathrm{tr}}(E)$ at $T = 300\,\mathrm{K}$. 
$\Gamma_{\mathrm{intra}}(E)$ and $\Gamma_{\mathrm{inter}}(E)$ have almost the same constant values of about $0.5\,\mathrm{meV}$. 
$2\Gamma_{\mathrm{tr}}(E)$ vanishes at $E=0 \,\mathrm{meV}$ owing to the screening effect, and approaches a constant value of about $1.5\,\mathrm{meV}$ as $E$ increases. 

\subsubsection{Alloy disorder scattering}
When there are alloy layers composed of ${\mathrm{A}}_x {\mathrm{B}}_{1-x} {\mathrm{C}}$, such as ${\mathrm{Al}}_x {\mathrm{Ga}}_{1-x} {\mathrm{As}}$ and ${\mathrm{In}}_x {\mathrm{Ga}}_{1-x} {\mathrm{As}}$, electrons are scattered by conduction band disorder. The scattering matrix element due to alloy disorder is given by \cite{Ando:1982-2,Bastard:1988}
\begin{eqnarray}
\left< |(m{\mathbf{k}}^{\prime}|H_1|n{\mathbf{k}})|^2 \right>
= \frac{a^3 (\delta E_c)^2 x(1-x)}{4} \int_{\mathrm{alloy}} dz 
  \left[ \zeta_m(z) \zeta_n(z) \right]^2,
\label{matrix element of alloy scattering}
\end{eqnarray}
where $a$ is the lattice constant and $\delta E_C$ is the difference in conduction band minima of crystals AC and BC (AlAs and GaAs in the case of ${\mathrm{Al}}_x {\mathrm{Ga}}_{1-x} {\mathrm{As}}$). 
Note that Eq. (\ref{matrix element of alloy scattering}) is independent of the scattering vector owing to the short-range nature of the scatterers. 
Therefore, we have 
\begin{eqnarray}
\Gamma_{\mathrm{intra}}^{\mathrm{(AD)}}(E) &=& 
 \frac{m^* a^3 (\delta E_c)^2 x(1-x)}{\pi \hbar^2} 
 \int_0^\pi d\theta \int_{\mathrm{alloy}} dz
 \left[ \frac{\zeta_0(z)^2}{S(q,T)} 
  - \zeta_1(z)^2 \right]^2,  \label{alloy-intra}\\
\Gamma_{\mathrm{inter}}^{\mathrm{(AD)}}(E) &=&
 \frac{m^* a^3 (\delta E_c)^2 x(1-x)}{\pi \hbar^2} 
 \int_0^\pi d\theta \int_{\mathrm{alloy}} dz
 \left[ \zeta_0(z)\zeta_1(z) \right]^2.  \label{alloy-inter}
\end{eqnarray}
$\Gamma_{\mathrm{inter}}(E)$ is independent of $E$, and $\Gamma_{\mathrm{intra}}(E)$ is also almost independent of it because $S(q,T) \sim 1$. 

On the other hand, the transport relaxation rate is given by \cite{Ando:1982-2,Bastard:1988}
\begin{eqnarray}
2\Gamma_{\mathrm{tr}}^{\mathrm{(AD)}}(E)
= \frac{2 m^* a^3 (\delta E_C)^2 x(1-x)}{\pi \hbar^2} \int_0^\pi d\theta
\frac{1 - \cos \theta}{\epsilon(q,T)^2} 
\int_{\mathrm{alloy}} dz \,\zeta_0(z)^4,   \label{alloy-tr}
\end{eqnarray}
which has an energy dependence due to the screening effect. 

Since alloy disorder scattering is due to $\delta E_C$, it can be regarded as a kind of roughness scattering. 
If one substitutes $V_0 = \delta E_C$, $\Delta^2 = a^2 x(1-x)/4$, and $\Lambda^2 = a^2/2\pi$ into Eqs. (\ref{ifr-intra}), (\ref{ifr-inter}), and (\ref{ifr-tr}), one can recognize that the alloy disorder scattering of Eqs. (\ref{alloy-intra})-(\ref{alloy-tr}) is expressed as the sum of the ``roughness scattering'' rates due to the alloy layer at position $z$.

Note here that $\Gamma_{\mathrm{intra}}(E)$, $\Gamma_{\mathrm{inter}}(E)$, and $2\Gamma_{\mathrm{tr}}(E)$ for alloy disorder scattering are similar in form to those for LA phonon scattering; thus $2\Gamma_{\mathrm{op}}(E)$ is comparable with $2\Gamma_{\mathrm{tr}}(E)$, as in LA phonon scattering.

\begin{figure}[bt]
\begin{center}
\includegraphics[width=.4\textwidth]{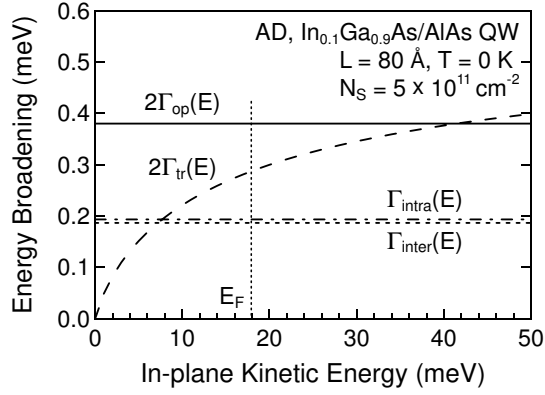}
\caption{$2\Gamma_{\mathrm{op}}(E)$, $\Gamma_{\mathrm{intra}}(E)$, $\Gamma_{\mathrm{inter}}(E)$, and $2\Gamma_{\mathrm{tr}}(E)$ due to alloy disorder (AD) scattering, plotted as functions of the in-plane kinetic energy $E$.}
\label{alloy-energy}
\end{center}
\end{figure}

Figure \ref{alloy-energy} shows $2\Gamma_{\mathrm{op}}(E)$, $\Gamma_{\mathrm{intra}}(E)$, $\Gamma_{\mathrm{inter}}(E)$, and $2\Gamma_{\mathrm{tr}}(E)$ in $\mathrm{In}_{0.1} \mathrm{Ga}_{0.9} \mathrm{As}/\mathrm{AlAs}$ QWs ($x=0.1$) at $T = 0\,\mathrm{K}$. 
We set the lattice constant and conduction band offset to approximately $a =5.66\,\mathrm{\AA}$ and $\delta E_C = 700\,\mathrm{meV}$, respectively. 
$\Gamma_{\mathrm{intra}}(E)$ and $\Gamma_{\mathrm{inter}}(E)$ have almost the same constant values of about $0.2\,\mathrm{meV}$. 
$2\Gamma_{\mathrm{tr}}(E)$ vanishes at $E=0 \,\mathrm{meV}$ owing to the screening effect, and approaches a constant value of about $0.57\,\mathrm{meV}$ as $E$ increases. 

\subsubsection{Ionized impurity scattering}
When dopant donors of Si are ionized, electrons supplied to QWs suffer from scattering by the Coulomb potential of the donors. 
The scattering matrix element due to an ionized impurity at position $Z$ is given by \cite{Ando:1985}
\begin{eqnarray}
(m{\mathbf{k}}^{\prime}|H_1|n{\mathbf{k}}) =
\frac{e^2}{2 \epsilon_0 \kappa_0 q} 
  \int dz\, \zeta_m(z) \zeta_n(z) \,e^{-q|z - Z|}.
\end{eqnarray}
Therefore, we have
\begin{eqnarray}
\Gamma_{\mathrm{intra}}^{\mathrm{(ION)}}(E) &=&
\frac{m^* e^4}{4\pi \epsilon_0{}^2 \kappa_0{}^2 \hbar^2}
  \int dZ N(Z) \int_0^\pi d\theta
  \left[ \frac{1}{q}
  \int dz \left\{ \frac{\zeta_0(z)^2}{S(q,T)}
  - \zeta_1(z)^2 \right\} e^{-q|z -Z|} \right]^2,  \\
\Gamma_{\mathrm{inter}}^{\mathrm{(ION)}}(E) &=&
\frac{m^* e^4}{4\pi \epsilon_0{}^2 \kappa_0{}^2 \hbar^2}
  \int dZ N(Z) \int_0^\pi d\theta
  \left[ \frac{1}{\tilde{q}}
  \int dz \,\zeta_0(z) \zeta_1(z) \, e^{-\tilde{q}|z -Z|} \right]^2,
\end{eqnarray}
where $N(Z)$ is the 3D impurity concentration at position $Z$. The transport relaxation rate, on the other hand, is given by \cite{Ando:1982,Hirakawa:1986}
\begin{eqnarray}
2\Gamma_{\mathrm{tr}}^{\mathrm{(ION)}}(E) 
&=& \frac{m^* e^4}{2\pi \epsilon_0{}^2 \kappa_0{}^2 \hbar^2}
  \int dZ N(Z) \int_0^\pi d\theta\, 
  \frac{1 - \cos\theta}{q^2 \epsilon(q,T)^2}
  \, \left[ \int \! dz \, \zeta_0(z)^2 \, e^{-q|z -Z|} \right]^2.
  \label{impurity-tr}
\end{eqnarray}

\begin{figure}[bt]
\begin{center}
\includegraphics[width=.4\textwidth]{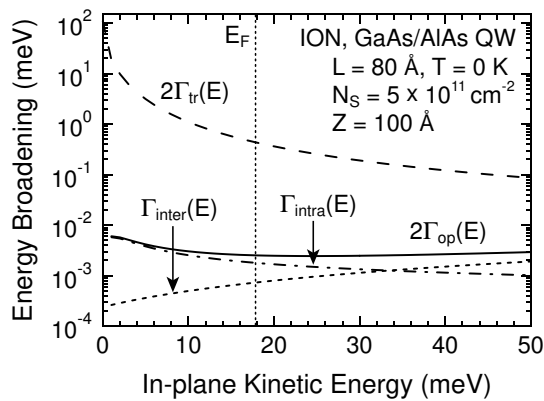}
\caption{$2\Gamma_{\mathrm{op}}(E)$, $\Gamma_{\mathrm{intra}}(E)$, $\Gamma_{\mathrm{inter}}(E)$, and $2\Gamma_{\mathrm{tr}}(E)$ due to ionized impurity (ION) scattering, plotted as functions of the in-plane kinetic energy $E$.}
\label{ion-energy}
\end{center}
\end{figure}

Figure \ref{ion-energy} shows $2\Gamma_{\mathrm{op}}(E)$, $\Gamma_{\mathrm{intra}}(E)$, $\Gamma_{\mathrm{inter}}(E)$, and $2\Gamma_{\mathrm{tr}}(E)$ in a $\delta$-doped GaAs/AlAs QW with $60\,\mathrm{\AA}$ spacer layers ($Z=100\,\mathrm{\AA}$) at $T=0\,\mathrm{K}$. 
First, $\Gamma_{\mathrm{intra}}(E)$ is much smaller than $2\Gamma_{\mathrm{tr}}(E)$, because the difference in intrasubband scattering matrix elements for the two subbands is small in ionized impurity scattering. 
Second, $\Gamma_{\mathrm{inter}}(E)$ is much smaller than $2\Gamma_{\mathrm{tr}}(E)$ owing to the large absolute value of scattering vector $\tilde{q}$. 
As a result, $2\Gamma_{\mathrm{op}}(E)$ is much smaller than $2\Gamma_{\mathrm{tr}}(E)$.

\section{Comparison with experimental results}
In this section, the absorption linewidth $2\Gamma_{\mathrm{op}}$ and transport energy broadening $2\Gamma_{\mathrm{tr}}$ are calculated for some GaAs (or InGaAs)/AlAs (or AlGaAs) QWs as functions of temperature, well width, alloy composition, and donor doping position. These results are compared with previously reported experimental data.

\subsection{Temperature dependence}
\begin{figure}[bt]
\begin{center}
\includegraphics[width=.4\textwidth]{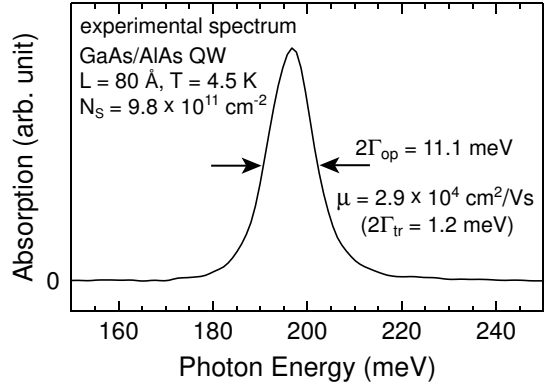}
\caption{Intersubband absorption spectrum observed at $4.5\,\mathrm{K}$ in a GaAs/AlAs single QW. The linewidth $2\Gamma_{\mathrm{op}}$ was $11.1\,\mathrm{meV}$ and the transport energy broadening $2\Gamma_{\mathrm{tr}}$ was $1.2\,\mathrm{meV}$, which was calculated from the mobility $\mu$ of $2.9 \times 10^4\,\mathrm{cm^2/Vs}$. 
}
\label{spectrum}
\end{center}
\end{figure}

\begin{figure}[bt]
\begin{center}
\includegraphics[width=.4\textwidth]{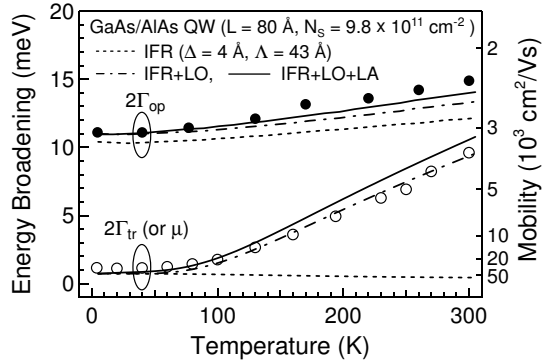}
\caption{Temperature dependence of the intersubband absorption linewidth $2\Gamma_{\mathrm{op}}$ and transport energy broadening $2\Gamma_{\mathrm{tr}}$ (or mobility $\mu$). Circles show experimental values, and lines show numerical results calculated by considering interface roughness (IFR), LO phonon, and LA phonon scattering.}
\label{Unuma T-dep}
\end{center}
\end{figure}

Experimental measurements of the temperature dependence of absorption linewidth are expected to clarify the effects of phonon scattering. 
We previously reported absorption linewidths in comparison with transport mobilities in a modulation-doped $\mathrm{GaAs/AlAs}$ single QW with a well width of $L=80\,\mathrm{\AA}$ and a sheet electron concentration of $N_S = 9.8 \times 10^{11} \,\mathrm{cm}^{-2}$, at temperatures ranging from 4.5 to $300 \,\mathrm{K}$ \cite{Unuma:2001}. 
The absorption spectrum observed at $4.5\,\mathrm{K}$ is shown in Fig. \ref{spectrum}. The low-temperature linewidth $2\Gamma_{\mathrm{op}}$ was $11.1\,\mathrm{meV}$ and the low-temperature transport broadening $2\Gamma_{\mathrm{tr}} = 2\hbar e/m_0^* \mu$ was $1.2\,\mathrm{meV}$, which was calculated from the mobility $\mu$ of $2.9 \times 10^4\,\mathrm{cm^2/Vs}$. Note that linewidth was one order of magnitude larger than transport broadening at low temperatures. The temperature dependences of linewidth and transport broadening are plotted in Fig. \ref{Unuma T-dep} by solid and open circles, respectively.

We performed the corresponding calculations of linewidth and transport broadening by considering interface roughness (IFR) scattering with $\Delta = 4\,\mathrm{\AA}$ and $\Lambda = 43\,\mathrm{\AA}$, LO phonon scattering, and LA phonon scattering. 
The contribution of alloy disorder scattering was absent because the GaAs QW had AlAs barriers, and the influence of ionized impurity scattering was sufficiently reduced by the spacer layers.

The calculated results for the linewidth $2\Gamma_{\mathrm{op}}$ versus temperature are also shown in Fig. \ref{Unuma T-dep} by dashed (IFR), dash-dotted (IFR+LO), and solid (IFR+LO+LA) curves, in comparison with the transport broadening $2\Gamma_{\mathrm{tr}}$. 
Additional width due to nonparabolicity is already included in these three curves, making small corrections compared with the contribution of interface roughness scattering, as seen in Fig. \ref{ifr-lambda}. 
At low temperatures, interface roughness scattering contributes $10.4\,\mathrm{meV}$ to linewidth, and LO-phonon spontaneous emission contributes $0.7\,\mathrm{meV}$. 
Phonon scattering processes become more active as temperature rises, and LO and LA phonon scattering contribute $1.8$ and $0.7\,\mathrm{meV}$, respectively, to linewidth at room temperature. 
These calculated results are in good agreement with the experimental data shown by solid circles. 
Note that the increase in dashed line (IFR) with increasing temperature is due to the nonparabolicity effect; the contribution of interface roughness scattering itself slightly decreases with increasing temperature, as expected from the energy dependence in Fig. \ref{ifr-energy}. 

For the transport broadening $2\Gamma_{\mathrm{tr}}$, interface roughness scattering makes a dominant contribution of $0.73\,\mathrm{meV}$ at low temperatures, which nearly explains the experimental value of $1.2\,\mathrm{meV}$. 
As already pointed out in the previous section, this value of $0.73\,\mathrm{meV}$ is an order of magnitude smaller than the contribution of $10.4\,\mathrm{meV}$ to linewidth. 
In the temperature range above $80\,\mathrm{K}$, the contribution of LO phonon scattering to transport broadening rapidly increases as temperature rises, and reaches a dominant value of $9.3\,\mathrm{meV}$ at $300\,\mathrm{K}$, as is well known. 
Such an effect of LO phonon scattering on transport broadening is very different from that on linewidth. 
The contribution of LA phonon scattering to transport broadening is $1.2\,\mathrm{meV}$, which is comparable with that to linewidth. 

As a result, linewidth and transport broadening have very different dependences on temperature. 
Similar behavior of linewidth versus temperature was also reported for $\mathrm{GaAs}/\mathrm{Al}_{0.3}\mathrm{Ga}_{0.7}\mathrm{As}$ QWs by Allmen \textit{et al.}\cite{Allmen:1988}.

\subsection{Well-width dependence}
Interface roughness scattering is expected to give absorption linewidth considerably strong well-width dependence. 
Campman \textit{et al.} reported low-temperature linewidths and mobilities in modulation-doped GaAs/Al$_{0.3}$Ga$_{0.7}$As QWs with $N_S \sim 6 \times 10^{11} \,\mathrm{cm^{-2}}$ for various well widths in the range $L = 75 - 110 \,\mathrm{\AA}$ \cite{Campman:1996}. 
Here, we calculate linewidth and transport broadening for the same structures. As scattering mechanisms, interface roughness scattering with $\Delta = 3\,\mathrm{\AA}$ and $\Lambda = 85\,\mathrm{\AA}$ and LO phonon scattering are included one by one.

\begin{figure}[bt]
\begin{center}
\includegraphics[width=.4\textwidth]{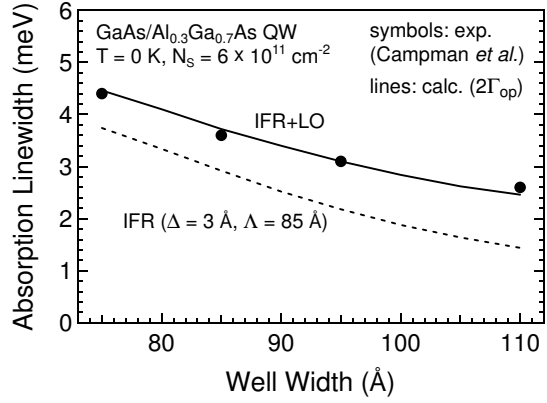}
\caption{Well-width dependence of intersubband absorption linewidth, calculated at $0\,\mathrm{K}$ by considering interface roughness (IFR) and LO phonon scattering. Solid circles show experimental results measured at low temperatures by Campman \textit{et al.} [9].}
\label{Campman-op L-dep}
\end{center}
\end{figure}

Figure \ref{Campman-op L-dep} shows the calculated results for low-temperature linewidth versus well width in the range $L = 75 - 110 \,\mathrm{\AA}$. 
First, the well-width dependence of linewidth due to interface roughness scattering, shown by the dashed curve (IFR), is not so strong for small $L$, because the confinement of the first excited state is weaker and thus $F_{11} = \partial E_1/\partial L$ is considerably smaller in $\mathrm{GaAs}/\mathrm{Al}_{0.3} \mathrm{Ga}_{0.7} \mathrm{As}$ QWs than in infinite-barrier QWs. 
Second, the contribution of LO phonon scattering slowly increases as QWs become wider, which makes the well-width dependence of linewidth slightly weaker. 
The solid curve (IFR+LO) is in good agreement with experimental results shown by solid circles \cite{Campman:1996}. 
If barriers are higher, as in GaAs/AlAs QWs, the first excited state is more strongly confined and the interface roughness scattering contributes much more to linewidth than LO phonon scattering does, which will lead to a much stronger well-width dependence of linewidth. 

\begin{figure}[bt]
\begin{center}
\includegraphics[width=.4\textwidth]{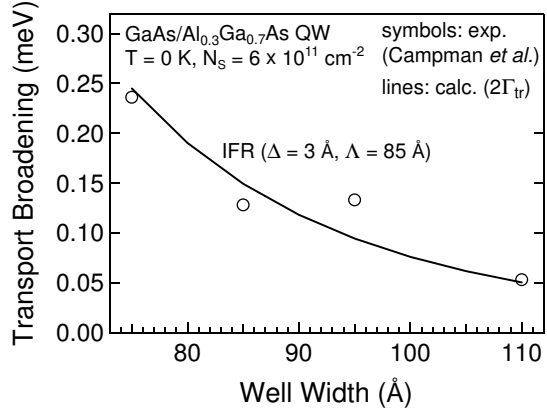}
\caption{Well-width dependence of transport energy broadening, calculated at $0\,\mathrm{K}$ by considering interface roughness (IFR) scattering. Open circles show experimental results measured at low temperatures by Campman \textit{et al.} [9].}
\label{Campman-tr L-dep}
\end{center}
\end{figure}

On the other hand, the well-width dependence of low-temperature transport broadening is shown in Fig. \ref{Campman-tr L-dep}. The transport broadening considered here is determined only by interface roughness scattering, because intrasubband LO-phonon emission and absorption are impossible at low temperatures. 
$F_{00}{}^2 = (\partial E_0/\partial L)^2$ in Eq. (\ref{ifr-tr}) is proportional to $L^{-6}$ in the infinite-barrier approximation, and this leads to a strong well-width dependence of transport broadening even in finite-barrier QWs. 
The calculated curve explains the experimental results plotted by open circles \cite{Campman:1996} very well.

\subsection{Alloy composition dependence}
Experimental measurements of the alloy composition dependence of linewidth are expected to show the effects of alloy disorder scattering. 
Campman \textit{et al.} reported low-temperature linewidths and mobilities in modulation-doped ${\mathrm{In}}_x {\mathrm{Ga}}_{1-x} {\mathrm{As/Al}}_{0.3} \mathrm{Ga}_{0.7} \mathrm{As}$ QWs with $L = 100\,\mathrm{\AA}$ and $N_S \sim 8 \times 10^{11} \,\mathrm{cm}^{-2}$ for various compositions in the range $x = 0 - 0.1$ \cite{Campman:1996}. We calculate linewidth and transport broadening for the same structures. 
As scattering mechanisms, interface roughness scattering with $\Delta = 3.5 \,\mathrm{\AA}$ and $\Lambda = 40\,\mathrm{\AA}$, LO phonon scattering, and alloy disorder (AD) scattering are included one by one.

\begin{figure}[bt]
\begin{center}
\includegraphics[width=.4\textwidth]{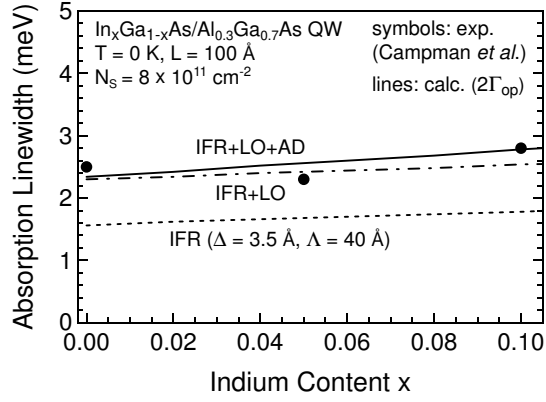}
\caption{Alloy composition dependence of intersubband absorption linewidth, calculated at $0\,\mathrm{K}$ by considering interface roughness (IFR), LO phonon, and alloy disorder (AD) scattering. Solid circles show experimental results measured at low temperatures by Campman \textit{et al.} [9].}
\label{Campman x-dep}
\end{center}
\end{figure}

Figure \ref{Campman x-dep} shows the calculated results for low-temperature linewidth versus alloy composition $x$ in the range $x = 0 - 0.1$. 
The contribution of interface roughness scattering is $1.6\,\mathrm{meV}$ at $x = 0$, and slowly increases as $x$ increases because QWs become deeper. 
LO phonon scattering contributes approximately $1\,\mathrm{meV}$ to linewidth, almost independently of $x$. 
Although the contribution of alloy disorder scattering is proportional to $x$ for small $x$, it is as small as $0.24\,\mathrm{meV}$ even at $x=0.1$. 
Our calculations explain the experimental observation of linewidth being insensitive to alloy composition, plotted by solid circles \cite{Campman:1996}. 

\begin{figure}[bt]
\begin{center}
\includegraphics[width=.4\textwidth]{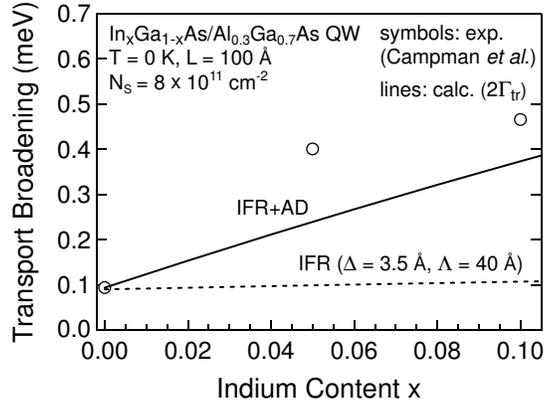}
\caption{Alloy composition dependence of transport energy broadening, calculated at $0\,\mathrm{K}$ by considering interface roughness (IFR) and alloy disorder (AD) scattering. Open circles show experimental results measured at low temperatures by Campman \textit{et al.} [9].}
\label{transport x-dep}
\end{center}
\end{figure}

On the other hand, transport broadening is shown in Fig. \ref{transport x-dep} as a function of $x$. 
Interface roughness scattering contributes $0.1\,\mathrm{meV}$ to transport broadening, while alloy disorder scattering makes the larger contribution of $0.27\,\mathrm{meV}$ at $x=0.1$; this shows that transport mobility drops remarkably as $x$ increases. 
Our calculations explain the experimental results plotted by open circles \cite{Campman:1996}. 
The small disagreement may be due to clustering in alloy layers, where the effective correlation length of alloy disorder in terms of roughness scattering may be larger than $a/\sqrt{2\pi}$ in actual samples grown by molecular beam epitaxy (MBE) or metalorganic chemical vapor deposition (MOCVD).

It should be noted that the one-order-of-magnitude different contributions of interface roughness scattering to linewidth and transport broadening are important in explaining their different behaviors versus alloy composition. 
Alloy disorder scattering in itself contributes fairly equally to linewidth and transport broadening, as mentioned in the previous section.

\subsection{Doping position dependence}
If the donor doping position is varied, then the contribution of ionized impurity scattering to linewidth should change. 
Dupont \textit{et al.} measured low-temperature linewidths in $\delta$-doped $\mathrm{GaAs/Al}_{0.25} \mathrm{Ga}_{0.75} \mathrm{As}$ QWs with $L = 76\,\mathrm{\AA}$ and $N_S \sim 1 \times 10^{12} \,\mathrm{cm}^{-2}$ for two different doping positions: $Z = 0$ and $112\,\mathrm{\AA}$ \cite{Dupont:1992}. 
We calculate linewidth and transport broadening for the same structures. 
As scattering mechanisms, interface roughness scattering with $\Delta = 5.66 \,\mathrm{\AA}$ (2 MLs) and $\Lambda = 70 \,\mathrm{\AA}$, LO phonon scattering, and ionized impurity (ION) scattering are included one by one.

\begin{figure}[h]
\begin{center}
\includegraphics[width=.4\textwidth]{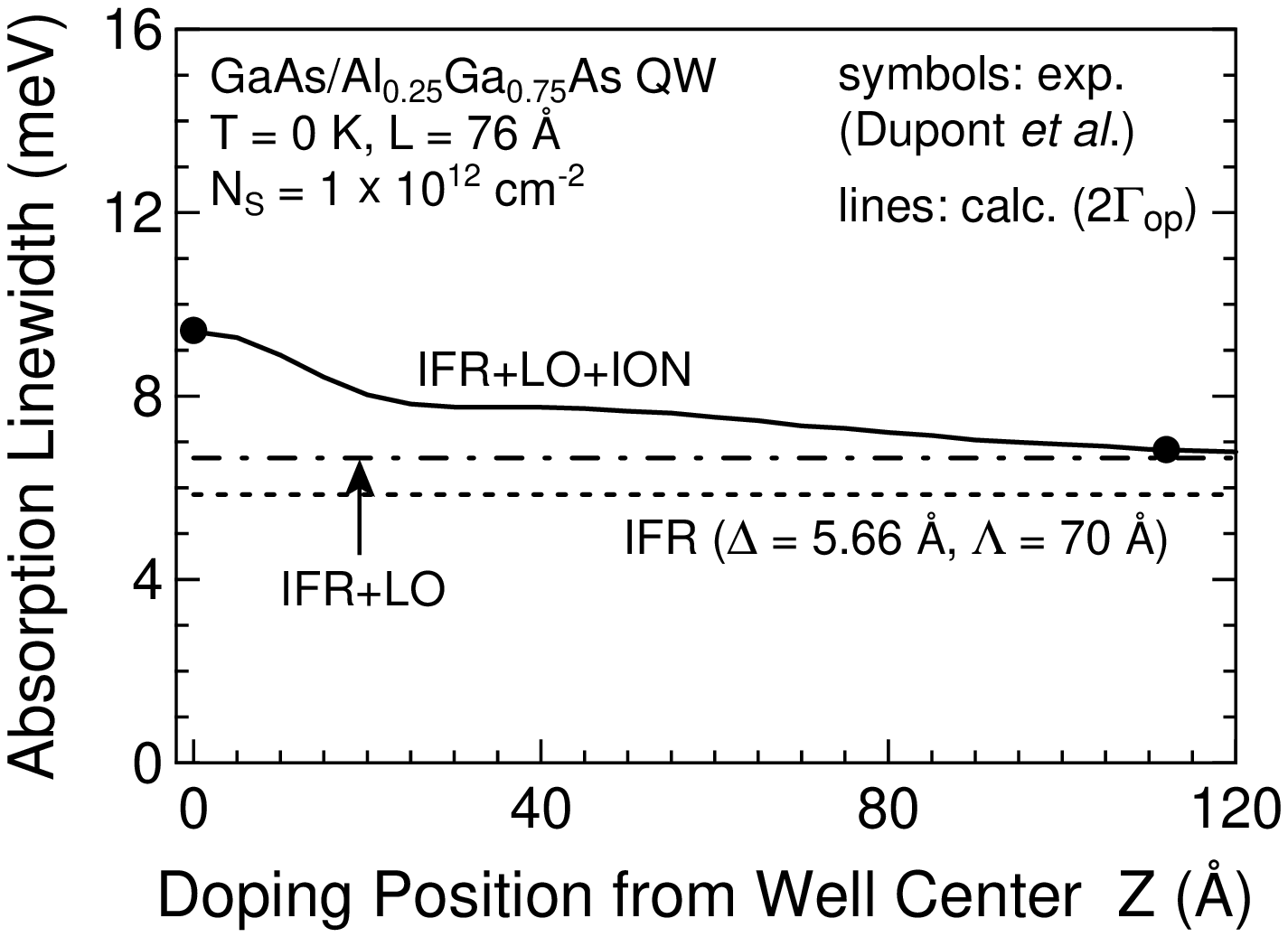}
\caption{Doping position dependence of intersubband absorption linewidth, calculated at $0\,\mathrm{K}$ by considering interface roughness (IFR), LO phonon, and ionized impurity (ION) scattering. Solid circles show experimental results measured at low temperatures by Dupont \textit{et al.} [10].}
\label{Dupont Z-dep}

\vspace{1cm}

\includegraphics[width=.4\textwidth]{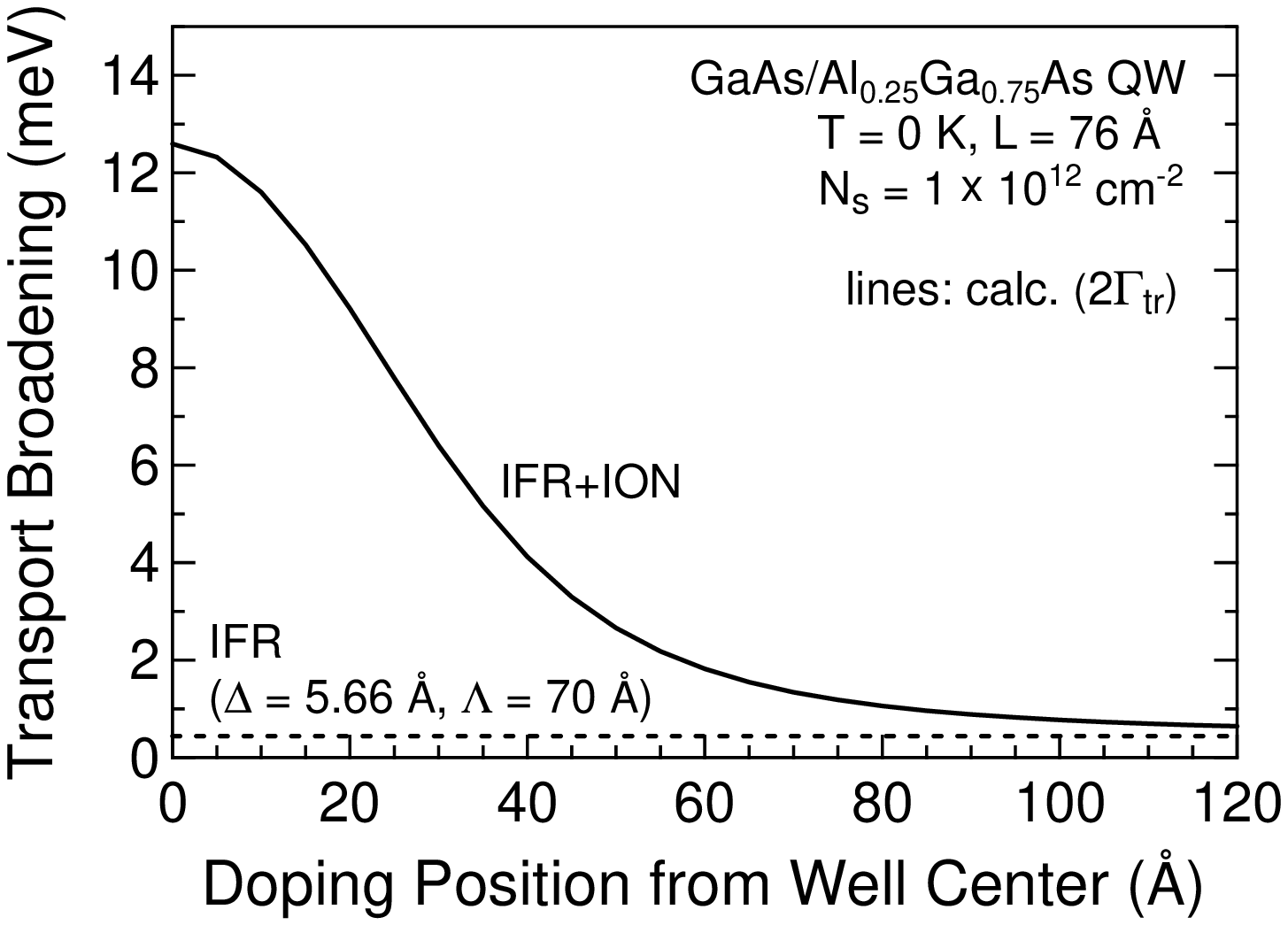}
\caption{Doping position dependence of transport energy broadening, calculated at $0\,\mathrm{K}$ by considering interface roughness (IFR) and ionized impurity (ION) scattering.}
\label{transport Z-dep}
\end{center}
\end{figure}

Figure \ref{Dupont Z-dep} shows the calculated results for low-temperature linewidth versus doping position $Z$ in the range $Z = 0 - 120\,\mathrm{\AA}$. 
Interface roughness scattering and LO phonon scattering contribute $5.8$ and $0.8\,\mathrm{meV}$ to linewidth, respectively. 
When donors are doped in barriers, at $Z = 100 \,\mathrm{\AA}$ for example, the contribution of ionized impurity scattering is as small as $0.3\,\mathrm{meV}$.  When donors are doped in QWs, at the center $Z = 0 \,\mathrm{\AA}$ for instance, the contribution of ionized impurity scattering is $2.8\,\mathrm{meV}$, which is smaller than that of interface roughness scattering. 
Our calculations explain the experimental results plotted by solid circles \cite{Dupont:1992}.

Note that the wave function $\zeta_1(z)$ of the first excited state penetrates largely into the low barriers in these narrow $\mathrm{GaAs/Al}_{0.25} \mathrm{Ga}_{0.75} \mathrm{As}$ QWs, so the effect of ionized impurity scattering is greatly enhanced even in barrier-doped QWs. 
If wave functions are more strongly confined, for example, as in the narrow GaAs/AlAs QWs used in our experiment, the contribution of ionized impurity scattering to linewidth is less than $0.1\,\mathrm{meV}$ for barrier-doping. 

On the other hand, low-temperature transport broadening is shown in Fig. \ref{transport Z-dep} as a function of $Z$. 
Interface roughness scattering contributes $0.44\,\mathrm{meV}$ to transport broadening, while ionized impurity scattering contributes $12.2 \,\mathrm{meV}$ at $Z=0\,\mathrm{\AA}$ and $0.33\,\mathrm{meV}$ at $Z=100\,\mathrm{\AA}$. 
Therefore, mobility greatly decreases when donors are doped in or near QWs; rather thick spacer layers, more than $150\,\mathrm{\AA}$ in this case, are necessary to completely remove the influence of ionized impurity scattering on mobility. 

\section{Summary}
We have formulated the microscopic energy-dependent relaxation rate $2\Gamma_{\mathrm{op}}(E)$ of intersubband optical transition in QWs due to scattering by interface roughness, LO phonons, LA phonons, alloy disorder, and ionized impurities, and have numerically calculated the absorption linewidth $2\Gamma_{\mathrm{op}}$ for GaAs-based QWs in comparison with the transport energy broadening $2\Gamma_{\mathrm{tr}}= 2\hbar e/m^* \mu$ related to the mobility $\mu$. 

The sensitivity of linewidth to interface roughness scattering is about one order of magnitude higher than that of transport broadening, because the contribution from the intrasubband scattering in the first excited subband is larger than that in the ground subband. 
This provides an essential insight for understanding experimental values for linewidth and the apparent lack of correlation between linewidth and mobility. 

The contribution of LO phonon scattering to linewidth is small, about $2\,\mathrm{meV}$ in narrow GaAs-based QWs even at room temperature, because the difference in intrasubband scattering matrix elements for the two subbands is small owing to the cancellation of form factors. 
In addition, intersubband LO-phonon spontaneous emission contributes approximately $1\,\mathrm{meV}$ to linewidth at low temperatures. 
Therefore, linewidth has a very weak temperature dependence, while mobility is greatly lowered by LO phonon scattering in the temperature range above $80\,\mathrm{K}$.

LA phonon scattering and alloy disorder scattering give matrix elements that are independent of scattering vectors, and lead to a linewidth comparable with transport broadening. 
The contribution of LA phonon scattering is, for example, about $1\,\mathrm{meV}$ at room temperature in narrow GaAs-based QWs, and this is small for linewidth compared with the contribution of interface roughness scattering. 
Alloy disorder scattering contributes, for instance, about $0.3\,\mathrm{meV}$ in ${\mathrm{In}}_x {\mathrm{Ga}}_{1-x} {\mathrm{As}}$ QWs with $x=0.1$. This is negligible for linewidth but predominant for transport broadening, causing a remarkable drop in mobility as $x$ increases.

Ionized impurity scattering contributes little to linewidth in modulation-doped QWs, because the difference in intrasubband scattering matrix elements for the two subbands is small. 
On the other hand, rather thick spacer layers, more than $150\,\mathrm{\AA}$ in narrow $\mathrm{GaAs/Al_{0.3}Ga_{0.7}As}$ QWs for example, are necessary to remove the influence of ionized impurity scattering on mobility.

\section{Acknowledgments}
We are grateful to Professor T. Ando for helpful discussions and showing us his unpublished formulation of intersubband optical transition. 
This work was partly supported by a Grant-in-Aid from the Ministry of Education, Culture, Sports, Science and Technology, Japan. 
One of us (T. U.) also thanks the Japan Society for the Promotion of Science for financial support.

\begin{table}[bt]
\caption{Material constants of GaAs.}
\begin{tabular}{ll}
\tableline
band gap of ${\mathrm{Al}}_x {\mathrm{Ga}}_{1-x} {\mathrm{As}}$ 
($x \le 0.45$) at $0\,\mathrm{K}$  &  ($1.519 + 1.247x) \,\mathrm{eV}$ \\
band gap of AlAs at $0\,\mathrm{K}$  &  $3.113\,\mathrm{eV}$     \\
conduction-band discontinuity ratio for GaAs/AlGaAs & $\sim 0.65$ \\
static dielectric constant   & $\kappa_0 = 12.91$ \\
optical dielectric constant  & $\kappa_\infty = 10.92$ \\
LO phonon energy  & $\hbar \omega_{\mathrm{LO}} = 36.5\,\mathrm{meV}$ \\
deformation potential constant & $D = 13.5 \,\mathrm{eV}$ \\
longitudinal elastic constant & $c_l = 1.44 \times 10^{11} \,\mathrm{N/m^2}$ \\
spin-orbit splitting & $0.341\,\mathrm{eV}$ \\
Kane energy & $22.7\,\mathrm{eV}$  \\
\tableline
\end{tabular}
\label{GaAs material constant}
\end{table}


\begin{references}
\bibitem{note1}
This paper does not deal with localized electrons that appear at lower electron concentrations.
\end{references}
\end{document}